%% file: relaxcs2-arxiv.tex
\documentclass[prl, twocolumn,superscriptaddress,showpacs,amsmath,amssymb]{revtex4}

\usepackage{graphicx}
\usepackage{bm}

\input{defs_thesis}

\begin{document}


\title{Electron spin relaxation of \natc~in \cstwo}


\author{John~J.~L.~Morton}
\email{john.morton@materials.ox.ac.uk}
\address{Department of
Materials, Oxford University, Oxford OX1 3PH, United Kingdom}
\address{Clarendon Laboratory, Department of Physics, Oxford
University, Oxford OX1 3PU, United Kingdom}

\author{Alexei~M.~Tyryshkin}
\address{Department of Electrical Engineering, Princeton
University, Princeton, NJ 08544, USA}

\author{Arzhang~Ardavan}
\address{Clarendon Laboratory,
Department of Physics, Oxford University, Oxford OX1 3PU, United
Kingdom}

\author{Kyriakos~Porfyrakis}
\address{Department of Materials, Oxford University, Oxford
OX1 3PH, United Kingdom}

\author{S.~A.~Lyon}
\address{Department of Electrical Engineering, Princeton
University, Princeton, NJ 08544, USA}

\author{G.~Andrew~D.~Briggs}
\address{Department of Materials, Oxford University, Oxford
OX1 3PH, United Kingdom}

\date{\today}

\begin{abstract}

We examine the temperature dependence of the electron spin
relaxation times of the molecules \natc~and \natcseventy~(which
comprise atomic nitrogen trapped within a carbon cage)~in liquid
\cstwo~solution. The results are inconsistent with the fluctuating
zero field splitting (ZFS) mechanism, which is commonly invoked to
explain electron spin relaxation for $S\ge1$ spins in liquid solution,
and is the mechanism postulated in the literature for these systems.
Instead, we find an Arrhenius temperature dependence for \natc,
indicating the spin relaxation is driven primarily by an Orbach
process. For the asymmetric \natcseventy~molecule, which has a
permanent ZFS, we resolve an additional relaxation mechanism caused
by the rapid reorientation of its ZFS.  We also report the longest
coherence time (\ttwo) ever observed for a molecular electron spin,
being 0.25~ms at 170K.

\end{abstract}


\pacs{76.30.-v,81.05.Tp}

\maketitle

\section{Introduction}

The encapsulation of atomic nitrogen within a fullerene shield has
provided a uniquely robust molecular electron spin~\cite{knapp97}.
Its unique relaxation properties have enabled the observations of a
novel type of electron spin echo envelope modulation
(ESEEM)~\cite{eseem05} and attracted attention as a potential
embodiment of a bit of quantum information~\cite{harneit}.

In high spin systems ($S\ge1$) in liquid solution, a fluctuating zero
field splitting (ZFS) has habitually been cited as the dominant
relaxation mechanism since transition metal ions were first studied
by EPR~\cite{mcgarvey, bloemmorgan}.  When relaxation in
\natc~(which has electron spin $S=3/2$) was first studied, it was
therefore natural to assume that the same ZFS mechanism
applied~\cite{knapp98}. However, to date there has been little
evidence to support this hypothesis. For example, no temperature
dependence has been reported for \natc~in solution; such a study is
critical in determining unambiguously which relaxation mechanisms
are relevant. Measurements have been reported in \cstwo ~and toluene
solutions~\cite{dietel99}; however, the analysis of these results
ignored the effects of magnetic nuclei in toluene, which we have
found to contribute significantly to the
relaxation~\cite{mortontolrelax}. Finally, the previous measurements
were performed using fullerene solutions that were sufficiently
concentrated for (\csixty$)_n$ aggregates to form, so it is
difficult to conclude which phase (liquid or solid) the reported
\tone/\ttwo~times correspond to~\cite{bokare03}. Consequently, the
favoured relaxation model of a zero-field splitting (ZFS)
fluctuation has little direct evidence to support it, and must be
critically re-evaluated.

In this letter we report relaxation times for both \natc~and
\natcseventy~in \cstwo~solution, which, conveniently, lacks nuclear
spins in the dominant isotopes of its constituents. We find that the
temperature dependence of the relaxation times is inconsistent with
the previously proposed ZFS mechanism, and suggest an alternate
Orbach relaxation mechanism. We extract an energy gap which matches
well the first excited vibrational state of the fullerene cage.

\section{Materials and Methods}
\begin{figure}[t]
\centerline {\includegraphics[width=3.2in]{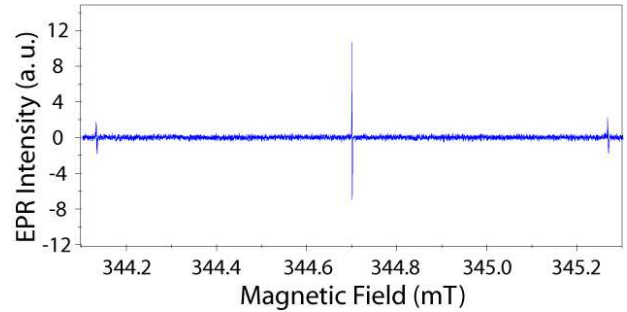}}
\caption{Continuous wave EPR spectrum of N@C$_{60}$ in CS$_{2}$ at
room temperature. Each line in the triplet signal is labeled with
the corresponding projection $M_I$ of the $^{14}$N nuclear spin.
Measurement parameters: microwave frequency, 9.67~GHz; microwave
power, 0.5~$\mu$W; modulation amplitude, 2~mG; modulation frequency,
1.6~kHz.}\label{cwEPR}
\end{figure}
High-purity endohedral N@C$_{60}$ was prepared~\cite{mito},
dissolved in CS$_{2}$ to a final fullerene concentration of
1-2$\cdot 10^{15}$/cm$^3$, freeze-pumped in three cycles to remove
oxygen, and finally sealed in a quartz EPR tube. The fullerene
concentration used ($\approx1~\mu$M) was well below the cluster
formation threshold~\cite{bokare03}. Samples were 0.7-1.4~cm long,
and contained approximately $5\cdot 10^{13}$ N@C$_{60}$ spins.
Pulsed EPR measurements were performed using an X-band Bruker
Elexsys580e spectrometer, equipped with a nitrogen-flow cryostat.
\ttwo~and \tone~times were measured using 2-pulse (Hahn) electron
spin echo (ESE) and inversion recovery experiments, respectively.
The $\pi/2$ and $\pi$ pulse durations were 56 and 112~ns
respectively. Phase cycling was used to eliminate the contribution
of unwanted free induction decay (FID) signals.

\Fig{cwEPR} shows the continuous-wave EPR spectrum of \natc~in
\cstwo~at room temperature. The spectrum is centered on the electron
g-factor $g=2.0036$ and comprises three narrow lines (linewidth
$<0.3~\mu$T) resulting from the hyperfine coupling to $^{14}$N
\cite{Murphy1996}. The relevant isotropic spin Hamiltonian (in
angular frequency units) is
\begin{equation}\label{Hamiltonian}
\mathcal{H}_0=\w_e S_z - \w_I I_z + a \!\cdot\! \vec{S} \!\cdot\!
\vec{I},
\end{equation}
where $\w_e=g\beta B_0/\hbar$ and $\w_I=g_I\beta_n B_0/\hbar$ are
the electron and $^{14}$N  nuclear Zeeman frequencies, $g$ and $g_I$
are the electron and nuclear g-factors, $\beta$ and $\beta_n$ are
the Bohr and nuclear magnetons, $\hbar$ is Planck's constant and
$B_0$ is the magnetic field applied along $z$-axis in the laboratory
frame. Each hyperfine line (marked in Fig.~\ref{cwEPR} with
$M_I=0$ and $\pm 1$) involves the three allowed electron spin
transitions $\Delta M_S=1$ within the $S=3/2$ multiplet. These
electron spin transitions remain degenerate for $M_I=0$ but split
into three lines for $M_I=\pm 1$. This additional splitting of
0.9~$\mu$T originates from the second order hyperfine corrections
and leads to a modulation of the electron spin echo decay~\cite{eseem05}.

\section{Relaxation of \natc~in \cstwo} \label{relaxcstwo}
\begin{figure}[t] \centerline
{\includegraphics[width=3.3in]{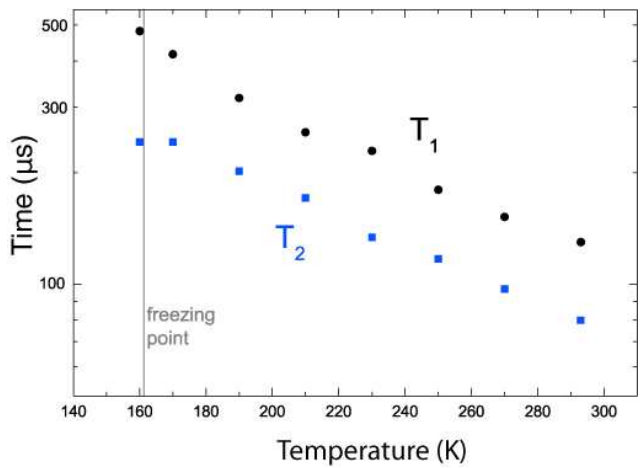}} \caption{Electron
spin relaxation times (\tone~and \ttwo) of \natc~in \cstwo, measured
using the central $M_I=0$ line.  The ratio \ttwo~$\approx(2/3)
$\tone~is maintained over the full temperature range for which the
solvent remains liquid.
 } \label{tempcs2}
\end{figure}

Spin relaxation times \tone~and \ttwo~for \natc~in \cstwo, measured
on the central $M_I=0$ hyperfine line, are shown on a logarithmic
scale in \Fig{tempcs2} for a range of temperatures (160K to 300K),
demonstrating an exponential temperature dependence and a roughly
constant ratio \ttwo~$\approx(2/3)$\tone~over the full temperature
range. This contrasts with previous findings which reported no
temperature dependence for \ttwo~\cite{harneit}. Below 160K, the
\cstwo~solvent freezes as a polycrystal, leaving regions of high
fullerene concentration around grain boundaries. This dramatically
increases the local spin concentration, and \ttwo~becomes extremely
short due to dipolar spin coupling (the so-called instantaneous
diffusion effect~\cite{klauder62,mims68,salikhov81}).

\begin{figure}[t] \centerline
{\includegraphics[width=3.3in]{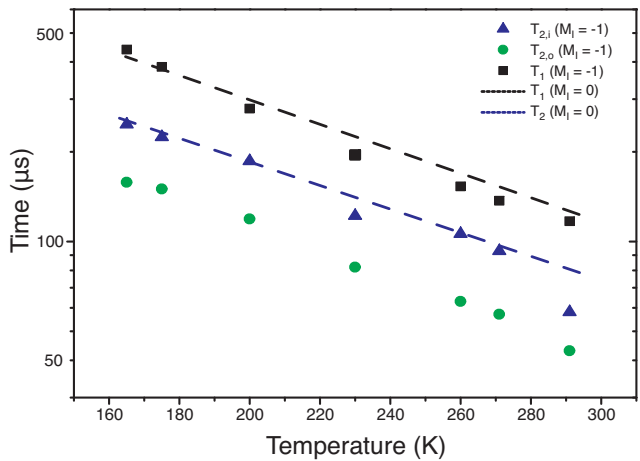}} \caption{Electron
spin relaxation times (\tone~and \ttwo) of \natc~in\cstwo, measured
using the high field $M_I=-1$ line. ESEEM is used to resolve the
individual decay rates of the inner and outer coherences (see
\Eq{eq:eseemt2}). Dashed curves show corresponding data taken for
the central  $M_I=0$ line, for comparison. } \label{t2t2c60}
\end{figure}

As this is an $S=3/2$ spin system, one might expect several
different relaxation times corresponding to the different $\Delta
M_S=1$ transitions. However, in the experiments presented in
\Fig{tempcs2}, all decays were well described by monoexponentials.
Given two similar exponential decays,
it is notoriously difficult to extract anything other than a single,
average decay constant from an exponential fit. Here, we take
advantage of a recently reported mechanism for electron spin echo
envelope modulation (ESEEM)~\cite{eseem05} to separate the
relaxation times for different electron transitions. This modulation
generates an echo intensity for transitions on the $M_I=\pm1$ lines
which varies as a function of the delay time, $\tau$, as
\begin{equation} \label{eq:eseem}
V_{M_I=\pm1}(\tau)= 2 + 3\cos2\delta\tau.
\end{equation}
The oscillating component arises from the `outer' coherences (from
the $M_S=\pm3/2:\pm1/2$ transitions), whilst the unmodulated
component arises from the `inner' coherences (from the
$M_S=+1/2:-1/2$ transition). If \ttwo~relaxation is included,
\Eq{eq:eseem} transforms to:
\begin{equation} \label{eq:eseemt2}
V_{M_I=\pm1}(\tau)= 2\exp{(-2\tau/\ttwoiq)} +
3\exp{(-2\tau/\ttwooq)} \cos2\delta\tau ,
\end{equation}
where \ttwoi~and \ttwoo~are the relaxation times of the `inner' and
`outer' coherences, respectively. Thus, by fitting to the modulated
ESEEM decay, the individual relaxation times \ttwoi~and \ttwoo~can
be extracted. \tone~and \ttwo~times measured for the high-field
($M_I=-1$) hyperfine line are shown in \Fig{t2t2c60}. \tone~was
measured in the standard way (inversion recovery), and so only one
(average) value was obtained.

The behaviour of \tone~appears identical for both central and
high-field lines, indicating that relaxation caused by the hyperfine
interaction with the nitrogen nuclear spin is negligible. The
\ttwoi~measured on the high-field $M_I=-1$ hyperfine line correlates
closely with the \ttwo~measured on the central $M_I=0$ line.
Remarkably, both of these \ttwo~times remain approximately 2/3 of
\tone~over the full temperature range studied. For the high-field
line, the ratio of \ttwoo~to \ttwoi~also stays constant at about
2/3. The fact that certain ratios between \tone, \ttwoi~and
\ttwoo~remain constant over a broad temperature range is a strong
indication that all of these relaxation times are limited by the
same mechanism. In the following section, we review different
relaxation mechanisms which might account for the observed
temperature dependence.

\subsection{ZFS fluctuations} \label{ZFS}
Spin relaxation is manifested in fluctuating terms in the spin
Hamiltonian and arises from fluctuating magnetic dipoles (either
nuclear or electronic), and other motions causing variations in the
interactions between the spin and its environment.  The trapping of
endohedral nitrogen in a high symmetry environment suppresses most of
the conventional spin relaxation mechanisms (zero-field splitting
(ZFS) interaction, anisotropic $g$ matrix, electron-nuclear dipolar
coupling and nuclear quadrupole interaction). Indeed, it has been
proposed that the dominant relaxation process arises from small
deviations from this ideal symmetric environment, caused by cage
deformations from collisions with solvent molecules~\cite{knapp97}.
For example, the modulation of the hyperfine interaction through
such collisions is a possible relaxation pathway. This was dismissed
in earlier reports on the basis that the expected $M_I$ dependence
of linewidth that this mechanism predicts is not
observed~\cite{knapp97}. However, as all linewidths are likely to be
instrumentally limited, this observation did not constitute a
rigorous confutation.

The mechanism favoured in the literature is that of a ZFS
fluctuation, again caused by deformation of the spherical
\csixty~cage through solvent collisions~\cite{knapp98}. Given the
concentrations of fullerene solution that were reported in these
earlier studies, a large amount of fullerene aggregation is
expected~\cite{bokare03} and so it is unlikely that the
\natc~molecules being studied had any direct contact with solvents.
Nevertheless, deformations of the cage, through whichever mechanism
(such as collisions with other \csixty~molecules in the cluster), will give
rise to some time-varying ZFS. Alternatively, ZFS fluctuations may
result from rotational tumbling in molecules that have a permanent
non-zero ZFS (such as in \natcseventy). In the case of a degenerate
$S=3/2$ system, a fluctuating ZFS term leads, in general, to two
different decoherence times~\cite{carrington},

\beq \label{zfst2t2dega}
\left(\ttwoiq\right)^{-1}=\frac{4}{5}D_{eff}^2\left[\frac{\tau_c}{1+\omega_e^2\tau_c^2}+\frac{\tau_c}{1+4\omega_e^2\tau_c^2}\right]
\eeq \beq \label{zfst2t2degb}
\left(\ttwooq\right)^{-1}=\frac{4}{5}D_{eff}^2\left[\tau_c+\frac{\tau_c}{1+\omega_e^2\tau_c^2}\right],
\eeq for the transitions that we refer to here as `inner' and
`outer' respectively. $D_{eff}^2=D^2+3E^2$,~$D$ and $E$ are the
coupling and rhombicity ZFS parameters, $\tau_c$ is the correlation
time of the fluctuations, and $\omega_e$ is the electron spin
transition frequency.

The predicted \tone~times arising from the same mechanism are:

\beq\label{zfst1t1dega}
\left(\toneiq\right)^{-1}=\frac{8}{5}D_{eff}^2\left[\frac{\tau_c}{1+\omega_e^2\tau_c^2}\right]
\eeq \beq \label{zfst2t1degb} \left(\toneoq
\right)^{-1}=\frac{8}{5}D_{eff}^2\left[\frac{\tau_c}{1+4\omega_e^2\tau_c^2}\right]
\eeq

The individual values of \tonei~and \toneo~cannot be resolved in a
simple inversion recovery experiment, and thus only their average
can be determined (with respective weights 2 and 3).

In the fast tumbling limit ($\omega_e \tau_c << 1$), the theory
predicts these two \tone~times to be identical, and equal to both
types of \ttwo,
contrary to our observed ratio of 2/3. Moving away from the
fast-tumbling limit, values for $D_{eff}$ and $\tau_c$ can be
derived given any values for \tone~and \ttwo. Since the ratio
between these times is dictated purely by $\tau_c$, the fact that
the ratios stay fixed implies $\tau_c$, the correlation time of the
ZFS fluctuations, stays fixed over the broad temperature range (160
to 300K). This would be surprising, as the viscosity of
\cstwo~changes by an order of magnitude over this temperature
range~\cite{kayelaby}. Thus, we conclude that the previously
suggested ZFS fluctuation mechanism cannot explain the observed
temperature dependence of \tone~and \ttwo, nor their mutual
correlation, and therefore seek alternative explanations for the
behaviour observed.


\subsection{Orbach relaxation process}
\begin{figure}[t] \centerline
{\includegraphics[width=3.1in]{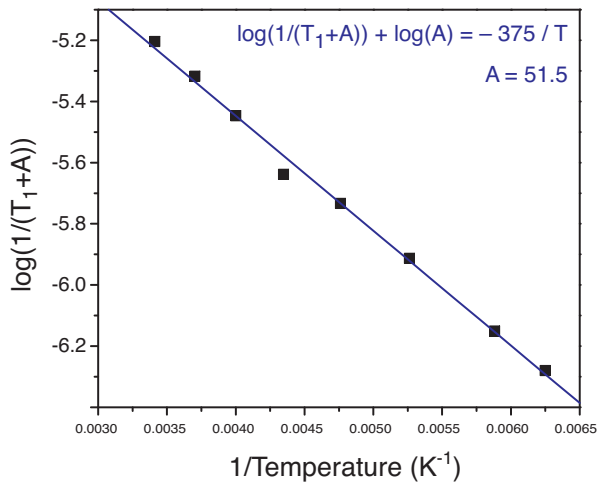}} \caption{The
temperature dependence of \tone~ of \natc~is linear in Arrhenius
coordinates, consistent with the Orbach relaxation mechanism. An
energy gap $\Delta = 32(1)~$meV $\equiv~375$K can be extracted.
Because we cannot make a low-temperature approximation in this case,
the standard Orbach plot of log(1/\tone) vs. 1/T must be adjusted to
include the constant of proportionality, $A$ (see Eq.~\ref{orbeq}).
The plot is then recursively fit to fine-tune $A$ and obtain the
slope, $\Delta/k$. \tone~is given in microseconds.}
\label{t1orbach}
\end{figure}
The temperature dependence of \tone~is well described by an Orbach
relaxation mechanism (see \Fig{t1orbach}). This is a two-phonon
relaxation process whose energies are resonant with a transition to
an excited electronic state (i.e.~a vibrational or orbital state
which lies outside of the space considered by the spin Hamiltonian).
The \tone~temperature dependence is dictated by the distribution of
phonon energies, and is of the form: \beq \label{orbeq} \toneq =
A~(e^{\Delta/kT}-1), \eeq where $\Delta$ is the energy gap to the
excited state and $A$ is some constant which involves terms
associated with spin-orbit coupling (and therefore with the ZFS,
\nfourteen~hyperfine coupling and g-tensor in the excited
state)~\cite{atkins72}. A fit to the data in \Fig{t1orbach} yields
$\Delta=32(1)$~meV. This is a close match to the energy of the first
vibrational mode of \csixty~(273~cm$^{-1}$, or 34~meV) which has
been theoretically calculated and observed by Raman spectroscopy of
\csixty~in \cstwo~solution at 300K~\cite{chase92, meilunas,vibc60},
indicating that this may be a vibrational spin-orbit Orbach
process~\cite{kivelson1,kivelson2}. This first excited vibrational
mode, termed $H_g(1)$, breaks the spherical symmetry of the
molecule, reducing it to axial. The small difference between
$\Delta$ observed here compared with that seen in the Raman
spectroscopy of \csixty~could be due to a shift in vibrational
energies due to the presence of the endohedral nitrogen atom.

The strong correlations observed in the temperature dependence of
\tone, \ttwoi~and \ttwoo~indicate that the \ttwo~times are also
limited by the Orbach mechanism. However, no detailed Orbach theory
has been developed for high-spin systems --- developing such a
theory lies beyond the scope of the current work.

\section{Relaxation of \natcseventy~in \cstwo}

\begin{figure}[t] \centerline
{\includegraphics[width=3.3in]{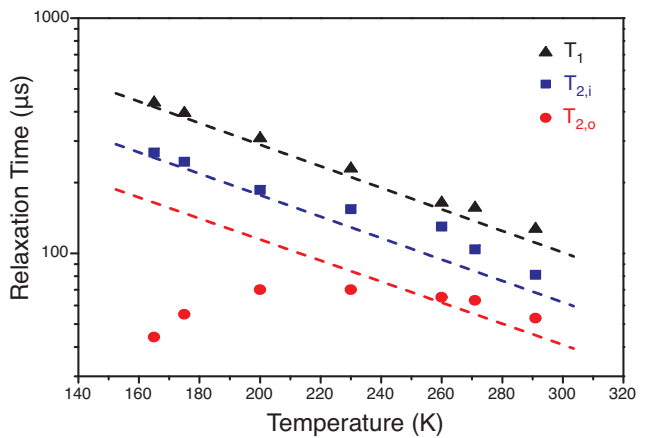}}
\caption{Temperature dependence of \tone~and \ttwo~times for
\natcseventy~in \cstwo. For comparison, dashed lines show linear
fits to the corresponding data for \natc~in \cstwo~(from
\Fig{t2t2c60}).} \label{relaxnc70}
\end{figure}

The Raman spectrum of \cseventy~is very similar to that of \csixty,
while its rugby ball shape provides a permanent non-zero ZFS to an
endohedral spin. \natcseventy~is therefore an ideal candidate to
further compare the mechanisms of a vibrational Orbach relaxation
with one induced by ZFS fluctuations (here, caused by molecular
rotations). Using the methods outlined above, we measured \ttwo~(for
both the inner and outer coherences) and \tone, shown in
\Fig{relaxnc70}.

\begin{figure}[t] \centerline
{\includegraphics[width=3.3in]{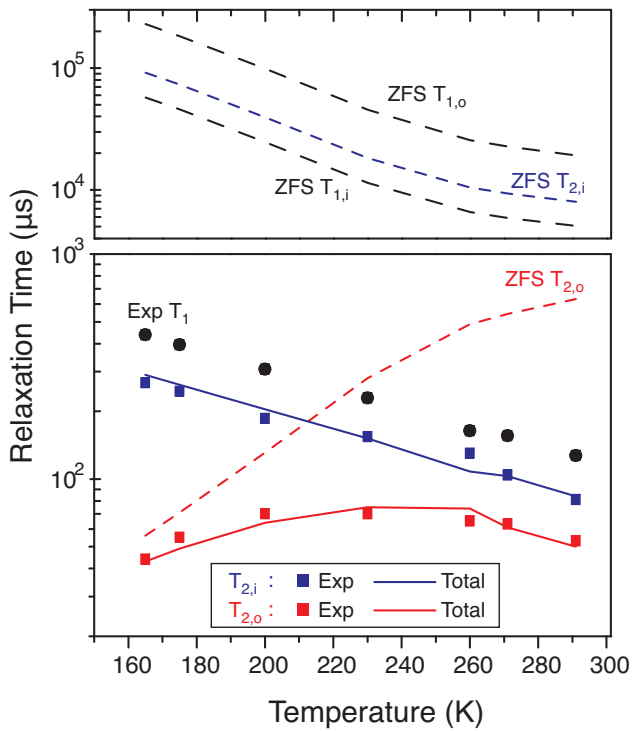}} \caption{Comparison
of \ttwo~times for \natcseventy~in \cstwo~solution with the model
described in the text. The curves labeled `ZFS' are derived from
Eqs.~\ref{zfst2t2dega} -- \ref{zfst2t1degb}. The `Total' fit to
\ttwoo~is achieved by combining the relaxation rate from the
fluctuating ZFS model, with an intrinsic decay taken to be 2/3 of
\ttwoi. The only free parameter in the model was a constant ZFS
parameter, $D$. The contribution of the ZFS model to \ttwoi~and both
\tone~is shown to be negligible (top panel).
} \label{nc70fit}
\end{figure}

The temperature dependence of \tone~is similar to that seen for
\natc~in \cstwo. The first excited vibrational mode of \cseventy~is
only about 1.7~meV lower in energy than the equivalent mode in
\csixty~\cite{dresselrev}. Consistent with this, the
\tone~temperature dependence seen for \natcseventy~is slightly
weaker than measured on the outer line of \natc, though the
difference falls within experimental error.

While \ttwoi~here bears a strong resemblance to that seen for \natc,
\ttwoo~for \natcseventy~shows a non-monotonic temperature
dependence, peaking around 230K. We now show that this behaviour can
be explained by the presence of the built-in ZFS in \natcseventy,
and by the change of rotational mobility of the molecule as the
temperature drops. An estimate of the built-in ZFS parameter in
\natcseventy~has been reported by aligning the molecules in a liquid
crystal, and was found to be $D=2.5$~MHz (0.8~G)~\cite{jakes02}.
However, due to the uncertainty in the order parameter ($O_{33}$),
this value should be considered as a lower limit of the true ZFS
parameter. At higher temperatures (i.e.~in the fast-tumbling regime)
this ZFS is averaged out sufficiently so that all relaxation times
are identical to those for \natc. However, upon cooling below 250K,
the viscosity of \cstwo~rises sharply~\cite{kayelaby}, thus slowing
the \natcseventy~tumbling rate and resulting in incomplete averaging
of the ZFS. We simulate this effect using
Equations~\ref{zfst2t2dega} and \ref{zfst2t2degb} and find that
while \ttwoo~is affected by this mechanism, both \ttwoi~and
\tone~are not.


In this simulation we assume that two relaxation mechanisms are
involved. One is the Orbach mechanism which produces the
correlations $\ttwoiq/\toneq~=~\ttwooq/\ttwoiq=2/3$ over the full
temperature range studied, as observed for \natc. The second is the
mechanism due to ZFS fluctuation, described above. The
Stokes-Einstein-Debye model, \beq \tau_r=\frac{4\pi \eta a^3}{3k T},
\eeq and experimental values for the viscosity of
\cstwo~\cite{kayelaby} are used to obtain the rotational correlation
time, $\tau_r$, as a function of temperature. The effective radius
of \cseventy~was taken to be
$5.4~\mathring{\mathrm{A}}$~\cite{khudiakov95}. The experimental
data were well fit by this model, using only one fitting parameter,
$D$ (given the axial symmetry of \cseventy, we assume $E=0$). The
result is shown in \Fig{nc70fit}, where the best-fit value for $D$
is 5.5~MHz (2~G). This value is large compared with estimates
described in the literature~\cite{jakes02}, however, it is
consistent with values for $D$ measured with other modifications of
\natc~(for example, $D$ was meausred in \natc O to be
2.4~G~\cite{oxidepaper}).

\Fig{nc70fit} also shows that the ZFS mechanism affects only \ttwoo,
and does not produce a noticeable effect on \ttwoi~and \tone.


\section{Conclusions}
In summary, we have reported the temperature dependences of electron
spin relaxation in nitrogen doped fullerenes, using ESEEM to resolve
the relaxation rates of different coherences of this $S=3/2$ spin.
Our findings are contradictory with the previously suggested
mechanism of a fluctuating ZFS, which is often assumed to be the
dominant mechanism in all high spin ($S\ge1$) systems. Instead, the
temperature dependences we observe are strongly suggestive of an
Orbach relaxation mechanism, via the first excited vibrational state
of the fullerene molecule. The study of electron spin relaxation in
the asymmetric \natcseventy~molecule permits us to distinguish this
Orbach relaxation mechanism from a fluctuating ZFS mechanism.
Additionally, the observation of a coherence time (\ttwo) in
\natc~of up to 0.25~ms, the longest for any molecular electron spin,
further emphasises the importance of this molecule for quantum
information processing. Such times allow in excess of 10$^4$ high
fidelity quantum gate operations to be performed~\cite{mortonbb1},
thus meeting the requirements for quantum error
correction~\cite{steane03}.

\section{Acknowledgements}
We acknowledge helpful discussions with Richard George, and thank
Wolfgang Harneit's group at F.U. Berlin for providing Nitrogen-doped
fullerenes, and John Dennis at QMUL, Martin Austwick and Gavin
Morley for the purification of \natc. We thank the Oxford-Princeton
Link fund for support. This research is part of the QIP IRC
www.qipirc.org (GR/S82176/01). GADB thanks EPSRC for a Professorial
Research Fellowship (GR/S15808/01). AA is supported by the Royal
Society. Work at Princeton was supported by the NSF International
Office through the Princeton MRSEC Grant No. DMR-0213706 and by the
ARO and ARDA under Contract No. DAAD19-02-1-0040.

\end{document}

%% file: defs_thesis.tex
\newcommand {\Eq}[1] {Eq.~\ref{#1}}

\newcommand {\Fig}[1] {Figure~\ref{#1}}

\newcommand{\beq}{\begin{equation}}
\newcommand{\eeq}{\end{equation}}

\newcommand{\natc}{N@C$_{60}$}
\newcommand{\natcseventy}{N@C$_{70}$}

\newcommand{\csixty}{C$_{60}$}
\newcommand{\cseventy}{C$_{70}$}

\newcommand{\cstwo}{CS$_2$}

\newcommand{\nfourteen}{$^{14}$N}

\newcommand{\beqa}{\begin{eqnarray}}
\newcommand{\eeqa}{\end{eqnarray}}
\newcommand{\w}{\omega}

\newcommand{\CCOMM}{Chem. Comm.}
\newcommand{\CPL}{Chem. Phys. Lett.}

\newcommand{\JCP}{J. Chem. Phys.}
\newcommand{\JAP}{J. Appl.. Phys.}
\newcommand{\JACS}{J. Am. Chem. Soc.}

\newcommand{\JMR}{J. Mag. Res.}
\newcommand{\JPC}{J. Phys. Chem.}

\newcommand{\MOLPHYS}{Mol. Phys.}

\newcommand{\PR}{Phys. Rev.}
\newcommand{\PRA}{Phys. Rev. A}

\newcommand{\PRL}{Phys. Rev. Lett.}

\newcommand{\ttwo}{T$_2$}
\newcommand{\ttwoi}{T$_{2,i}$}
\newcommand{\ttwoo}{T$_{2,o}$}
\newcommand{\tonei}{T$_{1,i}$}
\newcommand{\toneo}{T$_{1,o}$}
\newcommand{\tone}{T$_1$}

\newcommand{\toneq}{\rm{T}_1}
\newcommand{\ttwoiq}{\rm{T}_{2,\emph{i}}}
\newcommand{\ttwooq}{\rm{T}_{2,\emph{o}}}
\newcommand{\toneiq}{\rm{T}_{1,\emph{i}}}
\newcommand{\toneoq}{\rm{T}_{1,\emph{o}}}